\documentstyle[tighten,aps]{revtex}

\twocolumn

\newcommand{\be}{\begin{equation}}
\newcommand{\ee}{\end{equation}}
\newcommand{\ba}{\begin{eqnarray}}
\newcommand{\ea}{\end{eqnarray}}

\newcommand{\fr}[2]{\frac{#1}{#2}}
\newcommand{\non}{\nonumber}

\def\vec#1{{\mbox{\boldmath$#1$}}}

\newcommand{\p}{\mbox{$\vec{p}$}}

\newcommand{\r}{\mbox{$\vec{r}$}}

\newcommand{\lb}{\left (}
\newcommand{\rb}{\right )}
\newcommand{\la}{\left\langle}
\newcommand{\ra}{\right\rangle}
\newcommand{\ep}{\epsilon}
\newcommand{\vsig}{\mbox{$\vec{\sigma}$}}
\newcommand{\vSig}{\mbox{$\vec{\Sigma}$}}

\begin{document}

\draft

\title{
%
%
\[ \vspace{-2cm} \]
\noindent\hfill\hbox{\rm  SLAC-PUB-8652} \vskip 1pt
\noindent\hfill\hbox{\rm hep-ph/0010131} \vskip 10pt
%
%
${\cal O}(\alpha^3 \ln \alpha )$
corrections to muonium and positronium hyperfine splitting
}

\author{Kirill Melnikov\thanks{
e-mail:  melnikov@slac.stanford.edu}}
\address{Stanford Linear Accelerator Center, \\
Stanford University, Stanford, CA 94309}
\author{Alexander Yelkhovsky \thanks{
e-mail: yelkhovsky@inp.nsk.su}}
\address{Budker Institute for Nuclear Physics, \\
Novosibirsk 630090, Russia}
\maketitle

\begin{abstract}
We compute ${\cal O}(\alpha^3 \ln \alpha )$ relative corrections
to the ground state hyperfine splitting
of a QED two body bound state with different masses
of constituents.
The general result is then applied to muonium and
positronium. In particular, a new value of the muon
to electron mass ratio is derived from the muonium ground state
hyperfine splitting.
\end{abstract}


The  perturbative series for binding energies
of a QED bound state are non-analytic in the fine structure constant
and  the expansion contains  powers of $\alpha^{n_1} \ln^{n_2} \alpha$,
where $n_1$ and $n_2$ are some integer numbers. The appearance of
logarithms of $\alpha$  
is best explained by the fact that different scales, such
as the mass $m$, the   momentum $m \alpha$ and
the typical energy $m\alpha^2$  control  dynamics of the bound state.

In a recent paper \cite{pslog}
we have explained how the nonrelativistic Quantum
Electrodynamics (NRQED) regularized dimensionally can be efficiently
used to extract all $\ln \alpha$ corrections in a given order of
the expansion in $\alpha$ and applied this technique to compute
${\cal O}(\alpha^3 \ln \alpha)$ corrections to the decay rates
of para- and orthopositronium.  The purpose of this Letter
is to apply that methods  to the calculation of the  
${\cal O}(\alpha^3 \ln \alpha)$ corrections to the hyperfine 
splitting (hfs) of a general QED two body system in a ground state with an eye
on the hfs  in muonium and positronium.

For both, muonium and positronium,  there are good phenomenological reasons
to consider ${\cal O}(\alpha^3 \ln \alpha)$ contributions to the ground state
hfs.  The most precise measurement of this quantity
in positronium gives \cite{hughes}:
\be
\nu_{\rm exp}^{\rm Ps} = 203~389.10(74)~{\rm MHz},
\label{eksp}
\ee
while the theoretical prediction \cite{kp,
cmy},
which includes ${\cal O}(\alpha^3 \ln^2 \alpha)$ terms
computed in \cite{karsh},
is:
\be
\nu_{\rm th}^{\rm Ps} = 203~392.01(46)~{\rm MHz}.
\label{theory}
\ee
Obviously, the theoretical and experimental results differ from
each other by an uncomfortably large amount (given the claimed
accuracy of the two), which   indicates that
further study  of ${\cal O}(\alpha^3)$ corrections to the hfs
of the positronium ground state is warranted. The complete calculation
of ${\cal O}(\alpha^3)$ corrections is currently out of question
because of tremendous technical difficulties; nevertheless, the
${\cal O}(\alpha^3 \ln \alpha)$ corrections can be determined.

Before considering the case of two equal masses,
we  decided to study  a bound state of 
two particles with masses $m$ and $M$.
For $m=m_e$ and $M=m_\mu$ this corresponds to the bound
state of the electron and the anti-muon  called muonium.  
Our technique provides an excellent tool to extract the
${\cal O}(\alpha^3 \ln \alpha)$ corrections in this case keeping the
full mass dependence and this result can be used in two ways. First,
we will be able to check the correctness of our calculation
against the known results for ${\cal O}(\alpha^3 \ln \alpha)$
corrections obtained in an expansion in $m_e/m_\mu$. Second,
we will  derive some new results from our formula;
in particular, we will give the complete
${\cal O}(\alpha^3 m_e/m_\mu \ln \alpha)$
correction to the ground
state hfs in muonium
which is important
for the extraction of the muon to electron mass ratio.

The Letter is organized as follows.  We first review our method
of calculation (for more details we refer to \cite{pslog}).
We then consider the unequal mass case (muonium)
where  virtual annihilation is not allowed.
Later, we discuss phenomenological implications of our result for
muonium and derive the new value of the muon to electron mass ratio.
Finally, we describe how the calculation  should be modified
in order to accommodate the positronium case and its phenomenological consequences.

Our calculation is based on  dimensionally regularized
nonrelativistic QED
with $d=3-2\ep$ being the number of spatial dimensions.  
In the NRQED framework two different contributions to the final result
should be distinguished. The first one is the hard contribution,
which is sensitive to relativistic momenta only.
This contribution is not
capable to produce any non-analytic dependence on $\alpha$.
The second contribution
is the soft one. It is sensitive to  nonrelativistic scales
and for this reason can produce a non-analytic dependence on the
fine structure constant.  The main idea that permits a simple
extraction of the  logarithmic terms is the following.
In dimensionally regularized
NRQED, the matrix elements of the nonrelativistic operators are
the uniform functions of the fine structure constant. This implies
that, when written in proper units, the dependence on $\alpha$
can be scaled out of any matrix element.  We refer to our recent
paper \cite{pslog} for additional details on this approach; here
we remind the reader that the relative momentum $\p$, the relative
coordinate $\r$ and the binding energy $E$ scale as
$\p \to \gamma \p$, $\r \to \gamma^{-1} \r$, $E \to \gamma^2\mu^{-1} E$,
with the scaling parameter $\gamma = (\mu Z\alpha)^{1/(1+2\ep)}$
\cite{pslog}. Here $\mu = mM/(m+M)$ is the reduced mass of the bound state.
We also assign the charge $Z$ to the particle of mass $M$
to distinguish recoil and radiative recoil contributions, as it
is customary in bound state calculations. 
To illustrate how the scaling arguments help to
compute the $\ln \alpha$
corrections, let us consider the  matrix element of a nonrelativistic operator 
$O$ that delivers ${\cal O}(\alpha^3)$ correction
to the lowest order hfs kernel $V_{\rm Born}$ ($\vsig$ and $\vSig$ are the spin operators of the two particles),
\be
V_{\rm Born} = - \fr{Z\alpha}{mM}
                 \fr{[\sigma_i,\sigma_j]
                 [\Sigma_i,\Sigma_j]}{4d}\pi \delta(\r).
\ee
We consider relative correction to the hfs:
\be
\fr {\la \Psi |
O | \Psi \ra }{ \la \Psi | V_{\rm Born} | \Psi \ra }
= \frac {\alpha^3}{\pi} 
\lb \Delta_{
O} \ln (Z\alpha) + {\rm const} \rb.
\label{examp}
\ee
The operator $O$
is a function of  $\r$ and $\p$.
Performing the rescaling of all the quantities on the left hand side of Eq.(\ref{examp})
according to the rules given above, we extract the dependence on $\alpha$:
$$
\fr {\la \Psi |
O| \Psi \ra }{ \la \Psi | V_{\rm Born} | \Psi \ra }
=   \fr{\gamma^{3-n+j\epsilon}}{\mu^{l+k\ep}}
     \left. \fr {\la \Psi |
O | \Psi \ra }{
\la \Psi | V_{\rm Born} | \Psi \ra } \right|_{\gamma = 1},
$$
where 
$n=2,3$ is a power of $\alpha$
that explicitly enters
the operator $O$
and  $j,l,k$ are some integers. If the matrix
element is finite, we can  put $\ep=0$ and then
the relative correction to the hfs
is $\alpha^3$ times the $\alpha$-independent function
and hence no logarithms of $\alpha$ appear.
Therefore, after the rescaling, the {\it only} place where
$\ln \alpha$ can come from
is the expansion of the factor $\gamma^{3-n+j\epsilon}$ in powers
of $\ep$; this implies that, in order to generate the $\ln \alpha$
corrections, the nonrelativistic matrix elements should diverge and
only divergent pieces of the matrix elements are needed to
determine the  ${\cal O}(\ln \alpha)$ corrections.
Note also, that since the  hard ${\cal O}(\alpha^3)$ contributions
to the hfs
are not needed, it is
straightforward to keep the mass dependence exactly in our calculation.

Let us now list all the contributions to the hfs 
in muonium and positronium relevant at ${\cal O}(\alpha^3 \ln \alpha)$.
We begin with discussing the irreducible contributions, i.e. those that
arise as  average values of
some local operators. The first of those is the operator
that corresponds to a Taylor  expansion of ${\cal O}(\alpha)$
hard scale contributions in powers of the relative momenta of the  
bound state constituents up to ${\cal O}(\p^2)$. 
For both radiative and annihilation corrections
this contribution can be related to the divergences
in real radiation \cite{pslog}. For the recoil contributions 
things are more complicated and the easiest way to extract the 
${\cal O}(\alpha \p^2)$ piece of the hard scattering amplitudes 
is to actually expand the box diagrams to the required order.
This results in the following contribution to the muonium hfs:
\be
\Delta_{\rm real~rad} = Z^2\mu^2 
\left ( \frac {4}{3} \lb \fr{Z}{M} + \fr{1}{m} \rb^2
 + \frac {Z}{Mm}  \right ).
\label{realrad}
\ee

Other irreducible corrections at this order are produced due
to, loosely speaking,
the magnetic moment renormalization of the ${\cal O}(\alpha^2
)$
corrections to the hfs
and in many cases the result can be simply
obtained by generalizing the calculation of Ref.\cite{cmy}
to the unequal mass case. We then derive:
\ba
\Delta_{\rm rad~ret} &=&
       - Z^2(1+\xi Z^2)\fr{\mu^2}{m M},
\label{irrret} \\
\Delta_{\rm rad~1loop} &=& 
 Z^2 \mu \left ( \frac {2+\xi Z^2}{m} + \frac {1+2 \xi Z^2}{M}
       \right ).
\label{irr1loop}
\ea
for the contributions of the retardation
and the ``one-loop''
operators, respectively (see \cite{cmy} for the nomenclature).
Parameter $\xi=1$  distinguishes the contributions due to  anomalous
magnetic moment of the particle with the mass $M$.

Two additional irreducible contributions originate from  relativistic
corrections to the single Coulomb or magnetic exchange when we
account for the Pauli form factor in one of the vertices:
\ba
\Delta_{C} &=& - \fr{Z^2(1+\xi Z^2)}{4}
                 \fr{\mu^2}{m M},
\label{irrc}
\\
\Delta_{M} &=& -\fr{Z^2\mu^2}{4}
               \lb \fr{3+2\xi Z^2}{m^2} + \fr{2 +3 \xi Z^2}{M^2} \rb.
\label{irrm}
\ea

The last set of irreducible corrections can be loosely described as the 
effect of the retardation on all the relevant operators that generate non-radiative 
corrections at lower orders. This includes:
the third order retardation, the retardation in the one-loop operator,
the irreducible exchange of two magnetic photons, the retardation in 
the graph with magnetic seagull operator on one line, the graph with 
two magnetic-Coulomb seagull vertices on both lines and, finally, the 
graph with the magnetic seagull vertex on each line. 
The resulting  correction reads:
\be
\Delta_{\rm irr~ret} = \fr{22}{3} Z^3 
       \frac {\mu^2}{mM}.
\ee

Several reducible contributions appear in  the 
second order of time-independent perturbation theory.
The first of them is generated by the so-called double seagull
effective potential. This contribution is very similar to the
case of ${\cal O}(\alpha^3 \ln \alpha)$ corrections to the positronium
decay rate considered in \cite{pslog} and it can  be easily generalized
on the unequal mass case: 
\ba
\Delta_{\rm s}
 = \fr{Z^3 \mu^2}{m M} 
\lb
6 \ln (Z\alpha \mu^2) - \frac {2}{\ep}
+ \frac {20}{3} \left (\ln 2 -1 \right )
  \rb.
\label{seagres}
\ea

The reducible retardation correction can also be derived as a simple
generalization of the result in \cite{pslog}. The only difference is that 
the spin parts of the magnetic currents also give non-zero contribution to 
the hfs.  For  unequal masses, the result reads:
\ba
\Delta_{\rm ret} = \fr{Z^3 \mu^2}{m M} \lb
8\ln (Z\alpha \mu^2)
- \frac {8}{3\ep} 
+ \frac {64\ln 2}{3} - \fr{82}{3}
\rb.
\label{retres}
\ea

Also, the so-called ultrasoft contribution \cite{pslog}
 should be considered. We find:
$$
\Delta_{\rm us} =   - \fr{4 Z^2\mu^2}{3}
\lb \fr{Z}{M}  + \fr{1}{m} \rb^2
\lb
4\ln (Z\alpha\mu^{\fr{3}{2}}) - \frac {1}{\ep}
- \frac {5}{3}
\rb.
$$

For the last reducible contribution to the hfs,
the corresponding  nonrelativistic operator $O$ is
of the form $BGV_{\rm hl}+V_{\rm hl}GB$, where $G$ is the
reduced Green function of the Coulomb Hamiltonian,
$B$ is the Breit Hamiltonian and $V_{\rm hl}$ is the hard radiative
correction to $B$.  The Breit Hamiltonian in $d$ dimensions has been
derived in \cite{cmy} for $M=m$. For different masses, it reads:
\ba
B = && - \fr{ p^4 }{ 8 }
          \lb \fr{1}{m^3} + \fr{1}{M^3} \rb
           + \lb \fr{1}{2\mu^2} + \fr{ d-2 }{ mM } \rb
          \pi Z\alpha \delta(\r)
\nonumber \\
&&
        + \fr{ d-1 }{4} 
          \left\{ \fr{ p^2 }{ mM }, C \right\}
        - \fr{ \left[ [\vsig\nabla,\vsig]
          [\vSig\nabla,\vSig], C \right] }{ 16mM },
\label{breit}
\ea
where $C$ is the Coulomb potential in $d$ dimensions 
$
C(r) = - Z\alpha \Gamma(d/2-1)/(\pi^{d/2-1} r^{d-2} ).
$

The potential $V_{\rm hl}$  is 
the sum of four contributions
\cite{pineda}:
\be
V_{\rm hl} = V_{\rm ff} + V_{\rm magn} + V_{\rm box}
            + V_{\rm vp},
\label{vhl}
\ee
where the first one arises from the Coulomb photon exchange
with one of the vertices being either
the one-loop slope of the Dirac form factor
or the Pauli form factor,
\ba
V_{\rm ff} = \fr{2 Z \alpha^2}{3} &&\left[ \fr{1}{m^2}
                \lb - \fr{1}{\ep} + 2\ln m \rb
\right. \nonumber \\
             && \left. + \fr{Z^2}{M^2}
                \lb - \fr{1}{\ep} + 2\ln M -
\frac {3}{4} \left ( 1 - \xi \right ) \rb
                \right] \delta(\r),
\label{slope}
\ea
the second one is due to the one-loop anomalous magnetic moments,
\be
V_{\rm magn} = - (1+\xi Z^2)
          \fr{\alpha}{2\pi}
          \fr{ \left[ [\vsig\nabla,\vsig]
          [\vSig\nabla,\vSig], C \right] }{ 16mM },
\label{magn}
\ee
the third one comes from the hard one-loop box diagrams,
\ba
V_{\rm box} = \fr{(Z\alpha)^2}{ mM } &&\lb
              \fr{1}{\ep} - \ln (mM) - \fr{1}{3} \right.
              \non \\
              && \left.
              + \fr{M+m-2\mu(1+\vsig\vSig)}{M-m} \ln \fr{M}{m}
              \rb \delta(\r),
\label{box}
\ea
and the last one accounts for the one-loop vacuum
polarization  (hadronic vacuum polarization is not included):
\be
V_{\rm vp} = -\fr{4Z\alpha^2}{15}
            \lb \fr{1}{m^2} + \fr{Z^2}{M^2} \rb \delta(\r).
\label{vp}
\ee


The structure of $V_{\rm hl}$ is very similar to the structure of the Breit
Hamiltonian and so the corresponding calculation goes along the lines
of \cite{cmy}. On this way one recognizes that both $D$ wave and $S$ wave
contributions should be considered.

Since the $D$ wave part of the new perturbation (\ref{magn}) differs
from that of the original Breit perturbation (\ref{breit}) only by the
overall factor, we can  read off the $D$ wave contribution to the
${\cal O}(\alpha^3 \ln \alpha)$ correction to the hfs
from the corresponding ${\cal O}(\alpha^2)$ correction in positronium
(see \cite{cmy}):
\be
\Delta_D = - \fr{5}{12}(1+\xi Z^2)\fr{\mu^2}{mM}.
\label{dwave}
\ee

To find the contribution of the intermediate $S$ states, 
we first project both $B$ and $V_{\rm hl}$ on to the $S$ wave and 
then proceed along
the lines described in \cite{cmy}.  Since all the steps in this
calculation have their counterparts in the calculation described in detail
in \cite{cmy} and since the intermediate formulas for the different
mass case are lengthy, we refrain from presenting them here.

Summing up all the relevant contributions, we obtain the final result
for the ${\cal O}(\alpha^3 \ln \alpha)$ correction to the 
hfs of the ground state in the unequal mass case (we put $\xi=1$ below):
\ba
&& \Delta_{\rm tot} = 
\fr{Z^2\mu^2}{m^2}
\lb \fr{8}{3} \ln \fr{4m}{\mu Z\alpha} - \fr{281}{180} \rb
 - \fr{Z^2\mu^2}{mM} - \fr{Z^4\mu^2}{mM}
\non \\
&&
- \fr{Z^3\mu^2}{mM}
  \lb 2\ln \fr{mM}{\mu^2} + \fr{2\ln (Z\alpha)}{3}
- 20\ln 2 + \fr{101}{9} \rb
\label{gen}
\\
&& 
+ \fr{Z^3 \mu }{M-m}  \lb 5 + \fr{4\mu^2}{mM} \rb
\ln \fr{M}{m}  
+ \fr{Z^4 \mu^2}{M^2}
  \lb 
 \fr{8}{3} \ln \fr{4M}{\mu Z\alpha}
- \fr{281}{180} \rb
. \non
\ea

Eq.(\ref{gen}) is one of the principal results of this Letter. If we
identify $m$ with the electron mass and $M$ with the muon mass,
Eq.(\ref{gen}) provides the result for the ${\cal O}(\alpha^3 \ln \alpha)$
correction to  muonium hfs.  Muonium has been studied
extensively over the years and much is known about this system.
In particular, there are certain limits of Eq.(\ref{gen}) that
can be checked against known results.  To this end, it is instructive
to expand Eq.(\ref{gen}) in powers of $m/M$ up to the first non-trivial
order:
\ba
 \Delta^{\rm Mu} &&=  \fr{2Z^2}{3} \ln (Z\alpha)
 \lb - 4 +
 \frac {m_e}{m_{\mu}} \left ( 8 - Z \right ) \rb
\nonumber \\
&& + Z^2  \left \{
\frac {16}{3} \ln 2 - \frac {281}{180}
+ \frac {m_e}{m_\mu}
\left  [ -  Z^2 
- \frac {32}{3}\ln 2 + \frac {431}{90}
\right. \right. \non \\
&& \left. \left. 
~~~~~~~~ + Z \left ( 3 \ln \frac {m_\mu}{m_e} - \frac {101}{9} 
+ 20 \ln 2 \right )
\right ]
\right \}. 
\label{muexp}
\ea

The last equation shows  that our result, Eq.(\ref{gen}), correctly
reproduces the $\ln^2 (Z\alpha)$ terms \cite{laz,zwanz,karsh},
as well as
the $Z^2 \ln (Z\alpha)$ single logarithmic term \cite{laz,zwanz}
and the $(m_e /m_\mu)Z^3\ln (Z\alpha) \ln(m_\mu/m_e)$ 
term \cite{tomnio,karsh2}
which are all available in the literature. 

We now proceed to the discussion of what this result implies for the
phenomenology of the muonium hfs. We first note that
in this case Eq.(\ref{muexp}) can be used since
higher powers in the expansion in $m_e/m_\mu$ have a negligible impact.
Since the  $\ln (m_\mu/m_e)$ enhanced  part of the
${\cal O}(m_e/m_\mu \alpha^3 \ln \alpha)$ corrections has been properly taken
into account in a recent compilation of all  theoretical results
for muonium hfs \cite{mt},
we disregard it here. Setting $Z=1$, we then
obtain the ${\cal O}(m_e/m_\mu \alpha^3 \ln \alpha)$ hfs shift:
\be
\delta \nu^{\rm Mu} 
 = E_{\rm F}\frac { m_e}{m_\mu}
\frac {\alpha^3}{\pi} \left (
 \frac {28}{3}\ln 2  - \frac {223}{30} \right )\ln \alpha. 
\label{numshift}
\ee
Numerically, it evaluates to $0.013~{\rm kHz}$; this 
should be compared with a similar contribution
$\delta' \nu^{\rm Mu} = -0.265(64)
~{\rm kHz}$, originating  from the
incomplete calculation in \cite{tom}, that has been accounted for
in the theoretical value for muonium hfs in
\cite{mt}. 

The difference between the two numbers has
significant impact on the
electron to muon mass ratio determination. It is easy to see, that
it amounts to the relative shift of $6.2 \times 10^{-8}$
in this
ratio
(compare with the quoted relative theoretical uncertainty
$2.7\times 10^{-8}$ \cite{mt})
and, if we use the central value from \cite{mt}, Eq.(161),
the new result reads:
\be
\frac {m_\mu}{m_e} = 206.768~2784(30)(23).
\label{mrat}
\ee
Here the first error is related to the error in the theoretical prediction 
for the muonium hfs  and the second is the experimental one. 
The theoretical error in the hfs was estimated following
\cite{mt}; the only difference is that we estimated the uncalculated 
non-logarithmic ${\cal O}(m_e/m_\mu \alpha^3)$ recoil and radiative-recoil
corrections as half of their 
$\ln \alpha$ enhanced counterparts in Eq.(\ref{muexp}).
The uncertainty  in the muonium hfs 
due to uncalculated higher order corrections
we have obtained in this 
way is $0.07~{\rm kHz}$,  compared to $0.12~{\rm kHz}$ 
in \cite{mt}. 

For positronium, the calculation goes essentially unchanged, although
two facts have to be noticed. First, since
there is an annihilation contribution to the leading order hfs in positronium, the relative weight of different corrections
changes. Second, one has to take into account additional
annihilation contributions to the Breit and $V_{\rm hl}$ operators.
These annihilation operators read ($\vec{S}$ is the spin of positronium):
\be
B_{\rm ann} = \fr{\pi\alpha\vec{S}^2}{m^2} \delta(\r),
\ee
and
\be
V_{\rm ann} = \fr{4\alpha^2}{m^2}
              \left[ -1+\ln 2
              - \lb \fr{13}{18} + \fr{\ln 2}{2} \rb
              \vec{S}^2 \right] \delta(\r),
\ee
and they should be added to $B$ and $V_{\rm hl}$, respectively.
Finally, for obvious reasons,  one should disregard the ${\cal O}(1/M^2)$
contribution in Eq.(\ref{vp}). Proceeding along the lines described for
the unequal mass case, we  derive the
result for the ${\cal O}(m \alpha^7)$ hfs in positronium:
\be
\delta \nu^{\rm Ps} = \frac {7 m \alpha^7}{12\pi}
\ln \alpha \lb
- \frac {3}{2} \ln \alpha
+ \frac {68}{7} \ln 2 - \frac {62}{15} \rb.
\ee
Numerically, the new ${\cal O}(m \alpha^7 \ln \alpha)$ term
gives an additional shift of $-0.32~{\rm MHz}$ to the
theoretical value of the positronium ground state hfs, 
so that the theoretical prediction becomes:
\be
\nu_{\rm th}^{\rm Ps} = 203~391.69(16)~{\rm MHz}.
\label{theorynew}
\ee
The theoretical prediction moves closer to
the experimental result in Eq.(\ref{eksp}), however the difference is still
significant. Since the value of the new 
${\cal O}(m \alpha^7 \ln \alpha)$ contribution turns out to be roughly one 
third of the ${\cal O}(m \alpha^7 \ln^2 \alpha)$ one, the series 
look reasonably convergent.
For this reason we estimate the  nonlogarithmic ${\cal O}(\alpha^3)$
contribution to the positronium hfs as being one half of the
logarithmic one.  This is the origin of the 
uncertainty estimate in Eq.(\ref{theorynew}).  

In conclusion, we have computed ${\cal O}(\alpha^3 \ln \alpha)$
corrections to the hyperfine splitting of the general QED bound
state  keeping the full mass dependence. We then applied
this result to the hfs of the muonium and positronium.
The new value for the muon to electron mass ratio is extracted
from the ground state hyperfine splitting in muonium. As for 
the positronium ground state hfs, the computed correction slightly 
reduces the discrepancy between  theory and experiment. However, 
it is hard to imagine that higher order corrections can further 
significantly shift the  theoretical value.  In this circumstances,
one should perhaps start taking the discrepancy 
between the theory and experiment
in the positronium 
ground state hfs seriously.

{\it Acknowledgments:}
This work was supported in part by DOE under grant number
DE-AC03-76SF00515, and by the Russian  Foundation for
Basic Research grant 00-02-17646. We are indebted to 
R.~Hill for useful conversations and communicating to us
his results \cite{hill} on ${\cal O}(\alpha^3 \ln \alpha)$ corrections 
to the hfs  prior to publication. Discussions with 
A. Penin are gratefully acknowledged.

\end{document}